\documentclass[pre,aps,twocolumn,showpacs]{revtex4}

\usepackage{graphicx}
\usepackage{amsmath}
\usepackage{amsfonts}
\usepackage{amssymb}

\newcommand{\beqs}{\begin{eqnarray*}}
\newcommand{\eeqs}{\end{eqnarray*}}
\newcommand{\beq}{\begin{eqnarray}}
\newcommand{\eeq}{\end{eqnarray}}

\newcommand{\aS}{|S|}
\newcommand{\bit}{\begin{itemize}}
\newcommand{\eit}{\end{itemize}}

\newcommand{\al}{\alpha}
\newcommand{\p}{\partial}
\newcommand{\Voa}{V_{0,\al}}
\newcommand{\Woa}{W_{0,\al}}

\begin{document}
\title {Dewetting dynamics of stressed viscoelastic thin polymer films}
\author{Falko Ziebert}
\author{Elie Rapha\"el}
\affiliation{Laboratoire de Physico-Chimie Th\'eorique - UMR CNRS Gulliver 7083, ESPCI, 10 rue Vauquelin, F-75231 Paris, France}
\date{\today}
\begin{abstract}
Ultrathin polymer films that are produced e.g. by spin-coating
are believed to be stressed since polymers are 'frozen in' into out-of-equilibrium
configurations during this process. In the framework of a viscoelastic thin film model, 
we study the effects of lateral residual stresses on the dewetting dynamics of the film.
The temporal evolution of the height profiles and the velocity profiles inside the film
as well as the dissipation mechanisms are investigated in detail.
Both the shape of the profiles and the importance of frictional dissipation vs. viscous dissipation 
inside the film are found to change in the course of dewetting.
The interplay of the non-stationary profiles, the relaxing initial stress and  
changes in the dominance of the two dissipation mechanisms caused by nonlinear friction
with the substrate is responsible for the rich behavior of the system. 
In particular, our analysis sheds new light on the occurrence of the unexpected maximum in the rim width 
obtained recently in experiments on PS-PDMS systems.
\end{abstract}
\maketitle

\section{Introduction}
\label{Intro}

Thin polymer films are not only of obvious technological importance e.g. for coatings and lubrification
purposes but also have attracted recently the attention of physicists because of 
their rich dynamical behavior
\cite{Bankoff:1997,deGennes:2003,Bucknall:2004}.
On a non-wettable substrate, thin polymer films are unstable and start to dewet. 
For usual, purely viscous liquid films this has been known and studied 
already for some decades (see \cite{DeGennesRMP,LegerJoanny} for review articles on this subject), 
but for polymeric fluids the dynamics and phenomenology
is by far richer.
As thinner and thinner films became technologically feasible,
films produced with thicknesses smaller than the equilibrium size of a single polymer
have been studied \cite{Reiter:2001,Herminghaus:2001,GreenPRL,Reiter:2003.1}. 
These experiments revealed asymmetric rim shapes, 
several distinct regimes for the temporal evolution of the dewetting velocity
and a non-monotonous behavior of the width of the rim.

The asymmetric rim shapes could be reproduced by 
models assuming either a shear thinning fluid \cite{SaulnierPRL}
or a viscoplastic solid \cite{SharmaPRL}.  
Such nonlinearities in the mechanical properties
might well be present, but seem to be less essential for the formation of the rim 
than viscoelasticity and the friction with the substrate 
\cite{Herminghaus:2003,Vilmin_slippery,Fretigny}. This assertion is supported by
the absence of rim formation in
recent experiments on the dewetting of PS (polystyrene) films floating on a 
non-wettable liquid \cite{Fretigny}, where  
friction is avoided, while linear viscoelasticity could be clearly detected. 
In the experiments on supported films mentioned above \cite{Reiter:2001,Reiter:2003.1}, 
the silicium substrate is usually coated with PDMS (polydimethylsiloxane), leading to 
a polymer-polymer interface at the substrate inducing a strong slippage \cite{DeGennesCRAD79}. 
Indeed, in a viscoelastic model assuming slippage
and friction with the substrate,
the asymmetric shapes could be reproduced \cite{Vilmin_slippery}. 

Even more striking than the rim shapes has been the occurrence of a maximum 
in the width of the rim in the course of dewetting time
as observed by Reiter, Damman and coworkers \cite{Reiter:2003.1,Reiter:2005,Reiter:2007.2}
(for simple and for viscoelastic liquids one would expect a monotonous increase
of the size of the rim). 
This feature could be understood qualitatively using scaling arguments and 
modeled by assuming the friction with the substrate to be nonlinear
in the velocity (motivated by the polymer-polymer interface at the substrate) 
and by adding residual stresses inside the film that 
slowly relax during the dewetting process \cite{Reiter:2005,Vilmin_nlfric,Reiter:2007.2}.
These stresses are argued to originate from the spin-coating process,
where upon evaporation of the solvent 
polymers are 'frozen in' into out-of-equilibrium
configurations which induce an internal lateral stress \cite{Croll1979,Reiter:2005,Yang2006}. 
When the film is heated above the glass temperature
and starts to dewet, these residual stresses are able to 
relax, at least partially, and hence influence the dewetting dynamics. 
The existence of such stresses (for rather thick films where the occurrence of a shift
of the glass transition temperature could be excluded)
has been shown unambiguously \cite{Fretigny} and they have been estimated 
to be of the order of $10^5\,{\rm Pa}$.
The investigation of such stress-driven dewetting dynamics
is not only of importance for the stability, the mechanical properties and 
the dynamics of thin films,
but has even broader implications for the understanding of confined materials
and their glass transition 
\cite{Dalnoki2001,O_Tsui}. Indeed, the dynamics of the dewetting has been shown to be
severely influenced by the age of the sample, 
both on solid and liquid substrate \cite{Reiter:2005,Fretigny}. 

While there is nowadays a rather good understanding of the instability mechanisms leading to the 
film rupture \cite{HermJacobs:2005,Blosseyreview}, of the rim 
morphologies in the mature regime \cite{Herminghaus:2002},
and of the thin film equations 
for viscoelastic fluids \cite{Bankoff:1997,Thiele2001,Blossey:2005,Sharma_viscoelast},
the temporal evolution of the dewetting process is still rather unexplored.
In this work we investigate numerically the model introduced by Vilmin et al.
\cite{Vilmin_slippery}
with focus on the stress-driven dewetting dynamics.
Special interest is led on how the non-monotonous
behavior of the width of the rim arises - which could not be explained satisfactorily so far
by simple scaling laws
\cite{Vilmin_nlfric,Raphael:2006.1} which assume that friction is
the only relevant dissipation mechanism and that the film profiles
during dewetting are self-affine. 
Therefore, we also study the temporal evolution of the profiles, where the viscoelasticity 
shows to have important effects,
as well as the mutual importance of the two dissipation mechanisms - dissipation by friction 
with the substrate and viscous dissipation inside the film - in the course of time.

The model under investigation has been kept
simple and focuses on three main features: friction on slippery substrate, 
viscoelasticity and residual stress. 
It corresponds to the strong slip lubrication model of Ref.~\cite{muench2005} 
in the limit of zero Reynolds number
and without Laplace pressure.
Since we neglect the latter, the model only applies to the time window before
the round-up of the 'mature rim' is appreciable, which for the highly viscous polymer films
under consideration takes several reptation times. 
Also, instead of a hole geometry, we 
focus here on the simpler case of an edge geometry, i.e. a straight contact line. 
Then the system can be described by a one-dimensional model.
Experimentally, the dynamics in the hole and the edge geometry
have been compared in Refs.~\cite{Reiter:2003.1,ReiterEPJE} and 
a generalization of the model to the hole geometry is possible, see Ref.~\cite{Raphael:2006.1}.

This work is organized as follows: In section~\ref{Model}, we briefly 
review the model and the assumptions made
in its derivation. In section~\ref{anashort}, we give analytical expressions for 
the short-time behavior of the model, both
as a benchmark for the subsequent numerical work and 
to introduce the characteristic length and velocity scales.
Section~\ref{diml} reformulates the problem in dimensionless variables, the reduced parameters
are discussed and the numerical method is briefly described. 
Section~\ref{Num} shows numerical results for the physical observables: the dewetted length,
the velocity at the edge, the height of the rim at the edge and the width of the rim.
In the following section~\ref{Num_prof}, we have a closer look at the 
height, velocity and stress profiles.
Section~\ref{balance} is devoted to the energy balance in the course of dewetting
that is used to extract the two relevant dissipation mechanisms from the numerical results.
Finally we discuss the interplay between viscoelasticity, friction and
residual stress in section~\ref{Disc} and 
conclusions are presented in section~\ref{Concl}.

\section{Model}
\label{Model}

\begin{figure}
	\centering
	\includegraphics[width=0.45\textwidth]{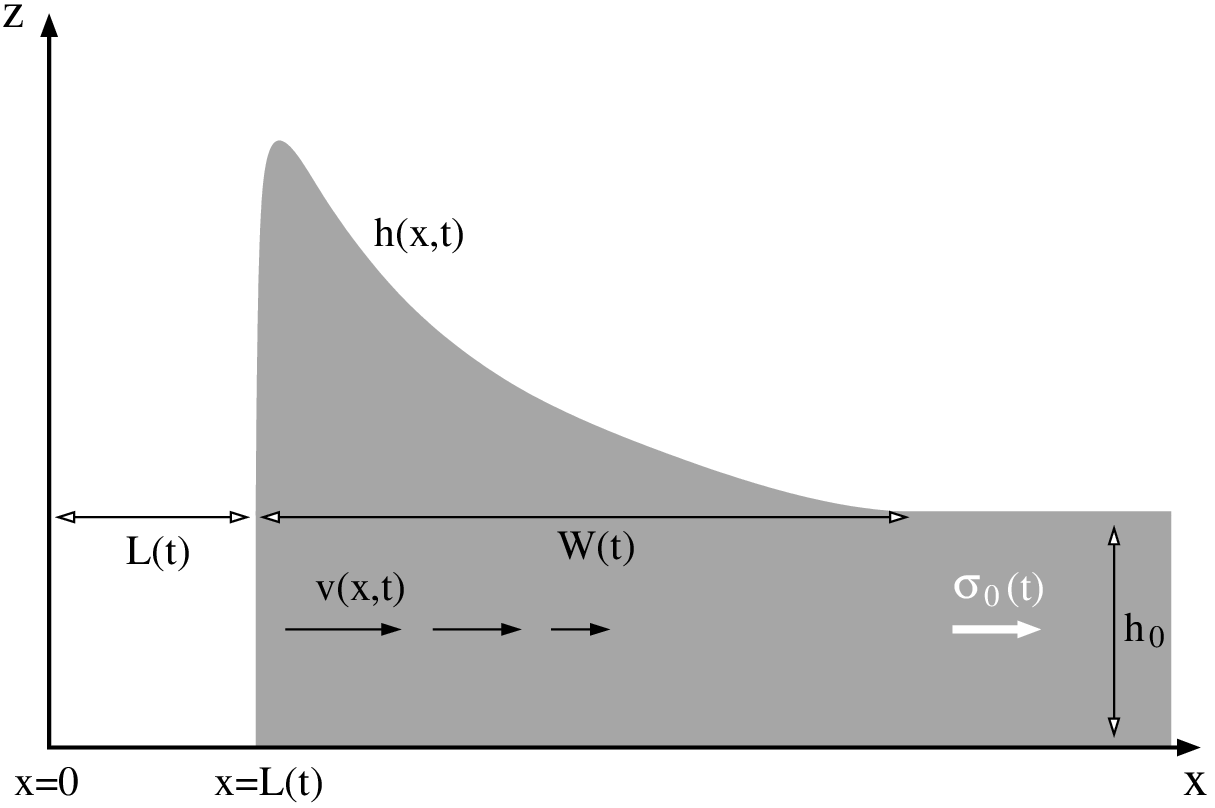}
	\caption{\label{sketch}
	Sketch of the thin film geometry. The height profile, $h(x,t)$, is represented in grey. 
	The dewetted length, $L(t)$,
	and the width of the rim, $W(t)$, are defined. The velocity, $v(x,t)$, is shown
	as black arrows; it decreases with the distance from the edge. 
	The residual stress, $\sigma_0(t)$, is indicated as a white arrow; it is assumed homogeneous
	in space, but relaxes in time (see text).
	}
\end{figure}

We consider dewetting from a straight edge to get an effective
one-dimensional description. Fig.~\ref{sketch} shows a sketch of this geometry.
Initially the film is flat with a vertical front at $x=0$ and 
the film extends infinitely in the direction $x>0$.
Since we are interested only in the dynamics of the rim for times smaller than
the time scale of the build-up of the mature rim, we neglect the Laplace pressure
arising from the film-air interface curvature. 
The film thickness $h(x,t)$ is assumed
to be small with respect to the hydrodynamic extrapolation length, or slippage length, $b=\eta/\zeta$,
where $\eta$ is the viscosity and $\zeta$ is the friction coefficient with the substrate 
\cite{DeGennesCRAD79}. 
This is the situation of interest for polymer films on polymer-covered substrates
and results in a plug flow in the film. 
In the spirit of a lubrication approximation, it is then sufficient to consider the horizontal
velocity $v(x,t)$, i.e. in the direction perpendicular to the dewetting front.
For details concerning the 
derivation and the lubrication approximation we refer to Ref.~\cite{Raphael:2006.1}.

In brief, 
the horizontal momentum equation (integrated over the thickness of the film)
is the balance of the frictional force at the film-substrate interface
and the divergence of the total stress inside the film.
We allow for either linear friction, $F_v=\zeta v$, or nonlinear friction,
\beq
F_v=\zeta \bar{v}^\al v^{1-\al}\,\,,\, 
\eeq
as motivated by recent experiments on polymer-polymer friction \cite{Leger}.
Here 
$\bar{v}$ is a characteristic velocity above which the friction is nonlinear
and $\al$ is the exponent characterizing the nonlinear behavior of the friction. 
If not specified otherwise we use
$\al=0.8$, a value obtained recently from experimental data on PS dewetting on PDMS-covered
substrates \cite{Vilmin_nlfric}, 
which is also in qualitative agreement with measurements on rubber-brush friction \cite{Leger}.
The divergence of the stresses in the film is
$F_s=\frac{\p\left(h\sigma\right)}{\p x}$.
Inertia is neglected due to the high viscosity of the polymer film, 
implying overdamped motion.
Momentum balance then leads to
\beq \label{mech}
\zeta \bar{v}^\al v^{1-\al}=\frac{\p\left(h\sigma\right)}{\p x}\,
\eeq
with $\al=0$ for linear friction and, in general, $0<\al<1$ for nonlinear friction.

To connect stresses and velocity gradients, a constitutive relation
has to be specified. We account for the viscoelasticity of the polymer film 
by using a standard Jeffrey model \cite{Bird},
\beq \label{const}
\sigma+\tau_1\p_t\sigma&=&G\tau_1\left[\left(\p_x v\right)+\tau_0\p_t\left(\p_x v\right)\right]\,.
\eeq
Here $G$ is the elastic modulus and $\tau_0$ and $\tau_1$, with $\tau_0\ll\tau_1$, 
are the two characteristic time scales.
By use of these one can define two viscosities \cite{comment}, 
$\eta_0=\tau_0 G=2\eta$  
and $\eta_1=\tau_1 G$.
The constitutive relation Eq.~(\ref{const})
describes viscous behavior with viscosity $\eta_0$ for times $t<\tau_0$,
elastic behavior for times $\tau_0<t<\tau_1$ and again
viscous behavior with viscosity $\eta_1$ for times $t>\tau_1$.
The latter, highly viscous behavior (since $\eta_1\gg\eta_0$) at long
times is supposed to reflect the slow dynamics by polymer reptation.
The time scale $\tau_1$ may be substantially smaller than the bulk relaxation time,
due to the thin film geometry \cite{Jones2005}, but has been recently proven to scale with molecular weight 
like the reptation time \cite{Reiter:2007.2}.
The short-time dynamics, $t<\tau_0$, involves intra-chain motions 
and is dominated by monomer friction resulting in a much lower viscosity.

The change in the height profile of the film
is given by volume conservation
\beq
\label{heq}
\p_{t} h&=&-\p_{x}\left(vh\right)\,.
\eeq
The position of the dewetting front is $L(t)$, starting from $L(t=0)=0$.
Its motion is governed by the velocity in the film at the edge, i.e. by $\p_t L=v(L)$.

Now we have to specify boundary conditions.
At the edge of the film, the height-integrated stress has to equal the driving force, hence
\beq
\label{bc}
h(L)\sigma(L)=-\aS\,.
\eeq
Here 
$S=\gamma_{sv}-\gamma_{sl}-\gamma$ is the so-called spreading parameter and
$\gamma_{sv}$, $\gamma_{sl}$ and $\gamma$ are the surface energies 
for solid-vapor, solid-liquid and liquid-vapor
interfaces respectively. In our case we consider a non-wettable substrate, i.e. $S<0$. 
Additionally we impose $v(x=\infty)=0$, assuming that the dewetting front is far
away from other fronts or holes and that the film is unperturbed far away.
As initial conditions we prescribe
$v(x,t=0)=0$ and $h(x,t=0)=h_0$, i.e. a quiescent film of height $h_0$.
The initial stress $\sigma(x,t=0)$ is either zero,
which we refer to as the case {\it without residual stress},
or given by a constant value throughout the film, $\sigma_0$,
which we refer to as the case {\it with residual stress}.

\section{Analytical short-time solution}
\label{anashort}

One now has to solve Eqs.~(\ref{mech}-\ref{heq}) with the initial and boundary conditions
specified above. In general, this has to be done numerically. 
For short times however, 
one can give insightful analytical expressions, due to the facts that 
the height profile is initially only slightly perturbed,
$|h(x)-h_0|\ll h_0$, and that the system is purely viscous at very short times, i.e.
$\sigma\simeq2\eta\p_x v$.

\subsection{Without residual stress}
\label{ana_nostress}

We first focus on the case without initial residual stress, i.e.
with $\sigma(x,t=0)=0$.
To leading order one has to solve
\beq
\zeta \bar{v}^\al v^{1-\al}=2\eta h_0\p_x^2 v\,.
\eeq
which in the case of linear friction, $\al=0$, yields 
an exponential velocity profile
\beq\label{vlin}
v(x,t)=V_0 \exp\left(-\frac{x-L}{\sqrt{2}W_0}\right)\quad\,\,{\rm for}\quad x>L\,.
\eeq
Here $L=L(t)$ denotes the dewetted distance, with $L(t=0)=0$.
$W_0$ is the characteristic rim width. 
By matching the boundary condition, Eq.~(\ref{bc}), one can identify
the characteristic initial velocity $V_0$. These two characteristic scales read 
\beq
\label{V0}
V_0=\frac{|S|}{\sqrt{2\eta\zeta h_0}}\,\,\,,\,\,\,
W_0=\sqrt{\frac{\eta h_0}{\zeta}}\,.
\eeq
The characteristic rim width can be rewritten 
as $W_0=\sqrt{h_0 b}$, where
$b$ is the extrapolation or slippage length, an expression that has been proposed some time ago
\cite{FBW:1994,FBW:1997}.

In the case of nonlinear friction, $\al\neq0$, 
a velocity profile of the following form is obtained: 
\beq\label{vprofalpha}
v(x,t)=V_{0,\al}\left(1-\frac{\al}{2}\frac{x-L}{\sqrt{2}W_{0,\al}}\right)^{\hspace{-1mm}\frac{2}{\al}}\,\,\, 
\eeq
for $0<x-L<\frac{2}{\al}\sqrt{2}\Woa$ and $v(x)=0$ elsewhere.
The corresponding characteristic scales now read
\beq
V_{0,\al}&=&\left(\left(\frac{2-\al}{2}\right)
\frac{V_0^2}{\bar{v}^\al}\right)^{\hspace{-1mm}\frac{1}{2-\al}}\quad\,\,{\rm and}\,\,\nonumber\\
\label{V0W0a}
W_{0,\al}&=&W_0\left(\left(\frac{2-\al}{2}\right)\frac{V_0^\al}{\bar{v}^\al}\right)^{\hspace{-1mm}\frac{1}{2-\al}}\,,
\eeq
with $V_0$ and $W_0$ as defined in Eq.~(\ref{V0}) above.

For the height profile at short times one simply has to solve
$\p_{t} h=-h_0\p_{x}v(x)$ which results in
\beq
h(x,t)&=&h_0\left[1+\frac{V_0 t}{\sqrt{2}W_0}\exp\left(-\frac{x-L}{\sqrt{2}W_0}\right)\right]
\,\,\,{\rm and}\,\,\nonumber\\
h(x,t)&=&\label{hprofalpha}h_0\left[1+\frac{V_{0,\al} t}{\sqrt{2}W_{0,\al}}
\left(1-\frac{\al}{2}\frac{x-L}{\sqrt{2}W_{0,\al}}\right)^{\hspace{-1mm}\frac{2-\al}{\al}} \right]\hspace{-1mm},\,\,\,\,\quad
\eeq
for linear and nonlinear friction, respectively.
In both cases, the rim build-up is initially linear with time, $h(L(t),t)\propto t$.
The stress inside the film is given by $\sigma(x)=2\eta\p_x v(x)$ and it can be easily verified
that  at the edge $h_0\sigma(L)=-\aS$ holds, and that the stress vanishes for $x\rightarrow\infty$. 

\subsection{Effect of residual stress}
\label{ana_stress}

We now consider the case of an initial residual stress, $\sigma(x,t=0)=\sigma_0$. 
Since the constitutive law is linear 
the total stress can be written
as $\sigma^{\rm tot}(x)=\sigma^{\rm v}(x)+\sigma_0$ (see sections~\ref{Num} and \ref{num_balance}
for a numerical validation).
Here $\sigma^{\rm v}$ represents the stresses in the absence of
the residual stress (the superscript ${\rm v}$ stands for {\it viscoelastic}). 
For the short-time response, we can again use $\sigma^{\rm v}=2\eta\p_x v$
and the stress balance at the edge reads
\beq\label{stotshort}
\sigma^{\rm tot}_{|L}=2\eta\p_x v(x)_{|L}+\sigma_0=-\frac{|S|}{h(L)}\,.
\eeq
Far away from the edge, there are no flows and additionally $\sigma(x=\infty)=\sigma_0$ must hold.
Using these boundary conditions, one gets the same formulas for
the velocity and height profiles, i.e. Eqs.~(\ref{vlin}), (\ref{vprofalpha}) and
Eqs.~(\ref{hprofalpha}) for linear and nonlinear friction respectively,
but with the substitution 
\beq\label{add_drive}
\aS\rightarrow\aS+h_0\sigma_0\,
\eeq
entering the characteristic velocity scale $V_0$,
and in the case of nonlinear friction entering both $V_{0,\al}$ and $W_{0,\al}$.

This clearly indicates that, as expected, the residual stress constitutes an
additional driving force for the dewetting.
For both linear and nonlinear friction, in the presence of a residual stress $\sigma_0$
the initial velocity of the dewetting
process is increased  
by a factor $(1+h_0\sigma_0/\aS)$ and 
$(1+h_0\sigma_0/\aS)^{\frac{2}{2-\al}}$, respectively.
The initial stress profiles in the presence of residual stress read
\beq\label{sprof_ana}
\sigma(x)&=&\sigma_0-\frac{\aS}{h_0}\left(1+\frac{h_0 \sigma_0}{\aS}\right)
\exp\left(-\frac{x-L}{\sqrt{2}W_0}\right)\quad\quad{\rm and}\,\,\nonumber\\
\sigma(x)&=&\sigma_0-\frac{\aS}{h_0}\left(1+\frac{h_0 \sigma_0}{\aS}\right)
\left(1-\frac{\al}{2}\frac{x-L}{\sqrt{2}W_{0,\al}}\right)^{\hspace{-1mm}\frac{2-\al}{\al}}\,\,\,\,\,
\eeq
for $\al=0$ and $0<\al<1$, respectively.

\section{Dimensionless equations, parameters and numerical method}
\label{diml}

To get access to the dynamics at longer times, one has to solve Eqs.~(\ref{mech}-\ref{heq}) numerically.
For this purpose, it is convenient to rescale these equations.
Keeping in mind the typical length and velocity scales obtained above,
we define the following dimensionless quantities
\beq
x'=\frac{x}{\Woa}\,\,,\,\,
v'=\frac{v}{\Voa}\,\,,\,\,
h'=\frac{h}{h_0}\,\,,\,\,
t'=\frac{t}{\tau_0}\,.
\eeq
At short times, $\sigma=2\eta\p_x v$ holds, so we rescale the stress as
$\sigma'=\sigma/\sigma^*$ with $\sigma^*=2\eta\frac{\Voa}{\Woa}=\sqrt{2}\frac{|S|}{h_0}$.

Thus we arrive at the dimensionless equations
\beq
\label{mech_n}
\left(\frac{2-\al}{2}\right) v'^{1-\al}&=& 2\p_{x'}\left(h'\sigma'\right)\,,\\
\label{const_n}
\sigma'+\tau_1'\p_{t'}\sigma'&=&\tau_1'
\left[\left(\p_{x'} v'\right)+\p_{t'}\left(\p_{x'} v'\right)\right]\,,\\
\label{heq_n}
\p_{t'} h'&=&-\beta\p_{x'}\left(v'h'\right)\,,
\eeq
with $\tau_1'=\tau_1/\tau_0$. We also have introduced an reduced parameter
\beq
\beta=\frac{\tau_0 V_0}{W_0}
=\sqrt{2}\frac{|S|}{Gh_0}\,,
\eeq
which is a dimensionless number quantifying
the coupling strength of the flow field to the height field, 
and which is proportional to $\aS/G$.

The rescaled boundary and initial conditions read
\beq\label{bound_ren}
h'(L)\sigma'(L)=-\frac{1}{\sqrt{2}}\,,
\eeq
$v'(x=\infty)=0$,
$v'(x',t'=0)=0$,
$h'(x',t'=0)=1$
and $\sigma'(x',t'=0)=0$ or $\sigma'_0$ respectively.
For the sake of simplicity, in what follows we will suppress the primes.

Remarkably, we are left with only three effective parameters: 
first, the friction exponent $\al$ that governs whether one deals with linear or nonlinear friction.
Second, the time scale $\tau_1'$ (in units of the
short time scale $\tau_0$) which governs the viscoelastic crossover from the elastic behavior 
of the film towards the highly viscous flow regime by reptation of chains in the film.
Finally, the parameter $\beta$,
describing the coupling between flow and height profiles.

We solved Eqs.~(\ref{mech_n})-(\ref{heq_n}) numerically, with the boundary and 
initial conditions specified above.
For this purpose, we discretized time and space. At every time step, we applied a shooting method
to solve Eqs.~(\ref{mech_n}) and (\ref{const_n}) simultaneously: i.e. at the edge we started 
with a trial velocity $v^*(L)$ 
(e.g. the one of the last time step) and the stress value prescribed by the boundary condition, 
Eq.~(\ref{bound_ren}), and evolved both equations
in space. This was done iteratively adjusting $v^*(L)$, until the 
boundary condition $v(x\hspace{-1mm}\gg\hspace{-1mm}1)=0$ was satisfied up to a prescribed tolerance.
Then we updated the height profile on a moving grid and proceeded to the next time step.

In this work we are not so much interested in the effects of parameter variations, but 
instead want to focus
on generic properties and the interplay between viscoelasticity, linear or nonlinear
friction, and the relaxation of residual stress.
Hence we use throughout the paper the parameter values
$\tau_1=100$, $\beta=0.2$ and $\al=0$ or $\al=0.8$ for linear and nonlinear friction, respectively. 
The initial residual stress was set either to $\sigma(t=0)=0$ or $1$. 

A difference of two orders of magnitude in the two time scales $\tau_0$ and $\tau_1$ 
is enough to separate the two viscous regimes 
and is a good compromise to avoid too time-consuming simulations. 
The actual value of $\tau_1/\tau_0$ is yet unknown
because the short time scale $\tau_0$ is hard to access experimentally. 
The value for the nonlinear friction exponent, $\al=0.8$, was used following 
the experimental results from Ref.~\cite{Vilmin_nlfric}.
The value of the parameter $\beta$ has been chosen for numerical convenience.
This parameter has only quantitative effects on the dynamics, 
and modest variations are noticeable only in the evolution of the height profile.
However, we use reduced variables - the elastic modulus $G$ and 
the wetting parameter $\aS$ that enter $\beta$ of course do have an important influence: 
the former is connected to both viscosities, 
and thus both enter the scales of the velocity and the width, see Eqs.~(\ref{V0},\ref{V0W0a}).

%
\section{Numerical results for physical observables}
\label{Num}
%

%
\begin{figure}
	\centering
	\includegraphics[width=0.45\textwidth]{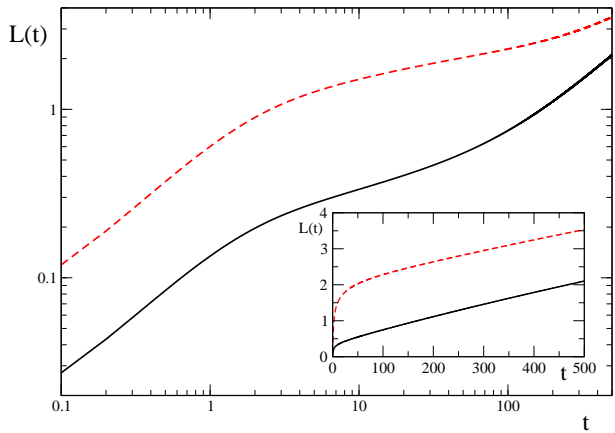}
	\caption{\label{ltsim}(Color online)
	The dewetted length, $L(t)$, is shown in double logarithmic scale.
	Nonlinear friction with $\al=0.8$ was used. 
	The black solid line shows the dependence in the absence of initial residual stress, while the
	red dashed curve has been obtained for $\sigma(t=0)=\sigma_0=1$.
	The inset shows the curves in linear scale.
	}
\end{figure}

In dewetting experiments, two observables are directly 
accessible by optical microscopy \cite{Reiter:2003.1}:
the dewetted length $L(t)$ and the width of the rim $W(t)$.
The velocity at the edge is then obtained by taking the temporal derivative
of the dewetted length, $\frac{dL}{dt}=V(t)=v(L(t),t)$.
The height of the film at the rim is not that easily accessible
(usually atomic force microscopy has to be used) 
and a full picture of the temporal evolution is hard to obtain.

\begin{figure}
	\centering
	\includegraphics[width=0.45\textwidth]{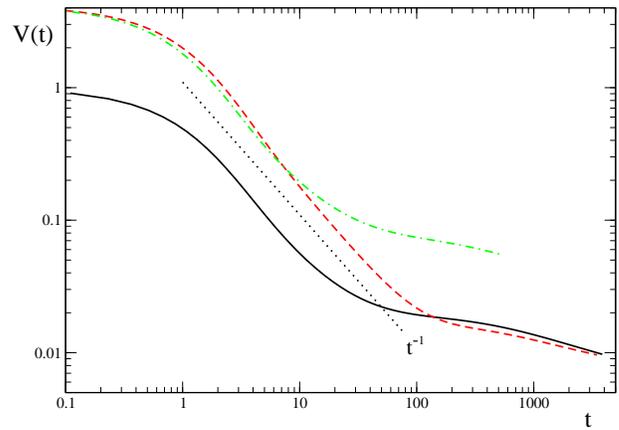}
	\caption{\label{vtsim}(Color online)
	The velocity at the edge, $V(t)=v(L)$, is shown 
	as a function of time in double logarithmic scale. Corresponding to Fig.~\ref{ltsim},
	the black solid line is in the absence of residual stress and the red dashed curve
	has been obtained for $\sigma(t=0)=\sigma_0=1$.
	The green dash-dotted curve shows a control simulation with a residual stress that
	is not allowed to relax (see text for details). 
	}
\end{figure}

In our simulations, we have direct access to these observables.
Fig.~\ref{ltsim} shows the dewetted length $L(t)$ in a double logarithmic scale
(see the inset for linear scale). 
Fig.~\ref{vtsim} displays the temporal evolution 
of the velocity at the edge, $V(t)=v(L)$, again in double logarithmic scale. 
In both figures, the black solid lines are the results without initial residual stress.
Since we have rescaled the velocity by $\Voa$, $V(t)$ starts at one. Once
the elastic regime is entered (for $t>\tau_0=1$) the velocity decreases
rapidly: the behavior is roughly $t^{-1}$ (see the dotted line in Fig.~\ref{vtsim}), 
as is discussed in detail in Ref.~\cite{Raphael:2006.1}.
Upon reaching the long time viscous regime ($t>\tau_1=100$), the velocity still decreases,
but only slowly. 
The same behavior can be deduced from Fig.~\ref{ltsim},
where the dewetted length is 
almost linear in time, actually slowly decreasing, for $t<1$ and $t>100$.

The red dashed lines in both figures have been obtained with an initial residual stress,
$\sigma(t=0)=\sigma_0=1$.
As expected, since the stress acts as 
an additional driving force (see Eq.~(\ref{add_drive})), 
the initial velocity is much higher in the presence  of
the residual stress. However, it also decays more rapidly due to the relaxation of this 
extra driving, until after a time of order $\tau_1$ both velocities are approximately the same.

\begin{figure}
	\centering
	\includegraphics[width=0.45\textwidth]{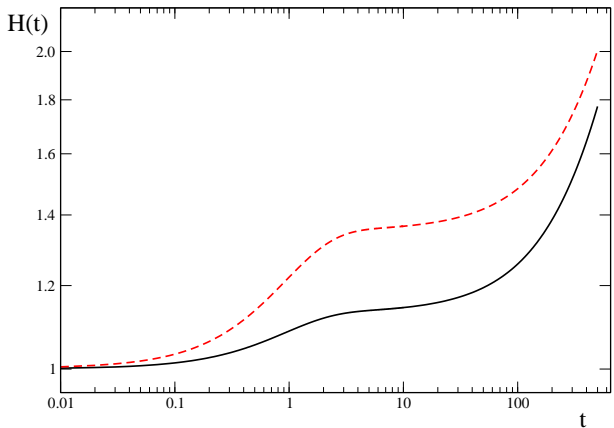}
	\caption{\label{htsim}(Color online)
	The height at the edge, $H(t)=h(L)$, is shown 
	as a function of time in double logarithmic scale. The black (solid) and red (dashed) curves 
	correspond to the respective curves displayed in the previous figures. They are 
	obtained with $\sigma(t=0)=0$ and $1$, respectively. One can clearly perceive the
	elastic plateau.
	}
\end{figure}

Figure~\ref{htsim} shows the temporal evolution of the height of the film at the edge, $H(t)=h(L(t),t)$.
The black (solid) and red (dashed) curves are 
obtained without initial stress and with $\sigma(t=0)=1$ 
and correspond to the respective velocities displayed in
Fig.~\ref{vtsim}. 
One can clearly discern the two viscous regimes at short and long times and the 
intermediate 'elastic plateau'.
The residual stress leads to more pronounced rims, i.e. higher values of $h(L)$. 
This could be expected from the short-time analysis,
 Eqs.~(\ref{hprofalpha}) and (\ref{add_drive}). It remains true for all times, because there
is no mechanism that could lead to a decrease of the height, even if the residual stress has relaxed.
This figure remains unchanged if linear friction ($\al=0$) is used instead. 

\begin{figure}
	\centering
	\includegraphics[width=0.45\textwidth]{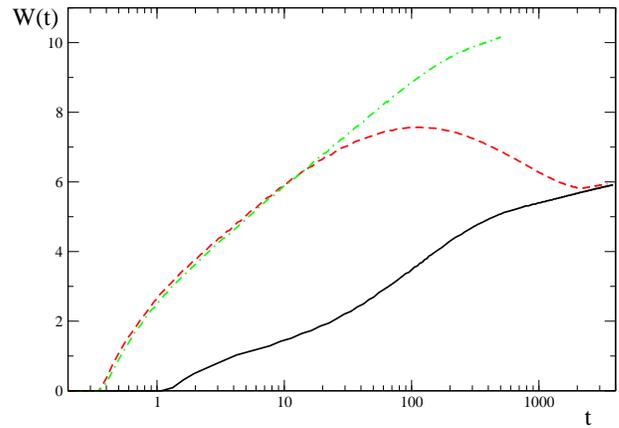}
	\caption{\label{wtsim}(Color online)
	The width of the rim, $W(t)$, as a function of $\log(t)$ 
	for the velocity evolutions shown in Fig.~\ref{vtsim}. 
	The black solid line, obtained without initial residual stress, and 
	the green dash-dotted line, obtained with a non-relaxing residual stress (see text for details), 
	display a monotonous growth of the width of the rim. 
	Only the red dashed curve, obtained with a residual stress that relaxes 
	(here with time scale $\tau_1$) has a maximum that is located close to $\tau_1$.
	}
\end{figure}

Figure~\ref{wtsim} shows the temporal evolution of the width of the rim, $W(t)$.
It has been obtained by introducing a cutoff as proposed previously \cite{Raphael:2006.1}: 
$W(t)$ is defined as the distance where $h(x,t)=(1+1/10)h_0$.
For usual liquids one would expect a monotonous increase in the width, 
even in the mature regime.
This is the case without initial residual stress (black solid curve) as well as with an 
initial stress that does not relax (green dash-dotted curve, see below). 
Only if the residual stress relaxes, here with the time scale $\tau_1$ given by the Jeffrey model, 
in the course of the dewetting process a maximum in the rim width is obtained 
as shown by the red dashed curve. 
This maximum is located close to the time scale $\tau_1$. 
The value of the maximum increases with increasing residual stress $\sigma_0$,
as has been studied previously \cite{Raphael:2006.1}.

We note that in this work 
we implement the residual stress as an initial bulk stress, $\sigma(t=0)=\sigma_0$,
see section~\ref{Model}.
This implies that the dynamics of this stress is governed
- as are the viscoelastic stresses - by the constitutive law, Eq.~(\ref{const_n}),
and hence it decays with the time scale $\tau_1$.
The green dash-dotted lines in Figs.~\ref{vtsim} and \ref{wtsim} have been obtained 
by a simulation where the residual stress was implemented differently:
As has been already proposed in section~\ref{ana_stress}, 
since the constitutive law is linear
one can separate the two stresses right from the start by writing
$\sigma^{\rm tot}(x)=\sigma^{\rm v}(x)+\sigma_0$. 
The residual stress, i.e. the contribution $\sigma_0$, then appears as the additional 
driving as described by Eq.~(\ref{add_drive}),
and additionally as a homogeneous contribution in the momentum equation, Eq.~(\ref{mech_n}).
One can use this separation, impose zero initial bulk stress and
consider $\sigma_0$ as a parameter that explicitly decays like 
$\sigma_0(t)=\sigma_0(t=0)e^{-t/\tau_1}$. Indeed, in doing so 
one exactly regains the results obtained by implementing an 
initial value for the stress in the bulk, i.e. the red dashed curves,
confirming that this separation in the stress is consistent.
In contrast, the green dash-dotted curves in Figs.~\ref{vtsim} and \ref{wtsim} have been obtained
by imposing $\sigma_0=1$ for all $t>0$. This represents the case where 
the residual stress does not relax (or only on time scales much larger than the time scale $\tau_1$).
Initially, the velocity at the edge and the width of the rim 
are the same for this case and for relaxing stresses, 
see the green and red curves,
hence again confirming the additional driving by the stress. 
However, there is no maximum in the rim width but a monotonous increase
if the residual stress does not relax.

At this point we can conclude that the model with residual stress
reproduces most of the features of the experiments in Refs.~\cite{Reiter:2001,Reiter:2003.1,Reiter:2005}, 
namely the high initial velocity, 
the fast decay of the dewetting velocity approximately like $v(L)\sim t^{-1}$, 
the elastic plateau in the height of the rim
and the maximum in the rim width.
It has been shown to be crucial that this residual stress relaxes. This 
relaxation dynamics then naturally influences the dynamics of the film.
To get a better understanding of the processes involved especially in the 
nonmonotonous behavior in the rim width,
in the following sections we focus on the temporal evolution of the profiles and
the dissipation mechanisms involved.

%
\section{Numerical results for the profiles}
\label{Num_prof}
%

For the system we are modeling, PS on PDMS-covered substrate,
the evolution of the height profiles of dewetting films has been studied in the
short-time regime by atomic force microscopy \cite{Reiter:2001},
revealing very asymmetric profiles. 
In principle, the height profile could be extracted also by 
investigating the interference patterns in optical micrographs.
Although velocity fields have been made visible e.g. in bursting of suspended 
soap films \cite{FBW:1995}, the velocity profiles are probably not 
obtainable by simple means for a highly viscoelastic polymer film. 

\begin{figure}
	\centering
	\includegraphics[width=0.45\textwidth]{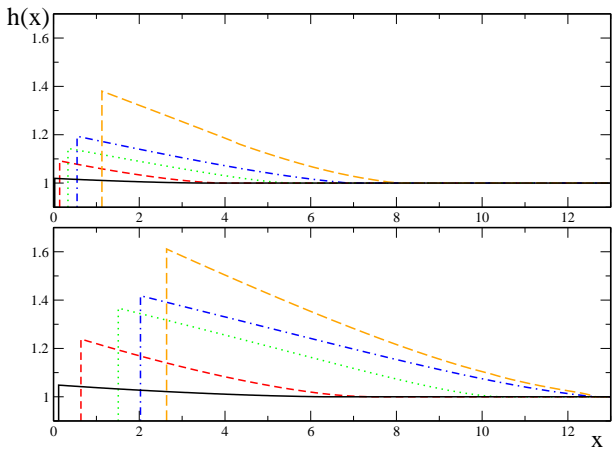}
	\caption{\label{hprof}(Color online)
	The height profiles $h(x)$ at successive times: 
	black (solid) curve $t=0.1$, red (dashed) $t=1$, green (dotted) $t=10$, blue (dash-dotted) 
	$t=50$, orange (long dashes) $t=200$.
	The upper panel shows the case without residual stress, the lower
	panel has been obtained with an initial residual stress $\sigma(t=0)=1$. 
	}
\end{figure}

In our simulations, we have direct access to the profiles and their
temporal evolution. Fig.~\ref{hprof} shows the height profiles of the film for successive times.
The upper panel displays the case without residual stress, while the lower panel
was obtained with $\sigma(t=0)=1$. At short times, we get indeed the profiles
as described by Eq.~(\ref{hprofalpha}), in the case with residual stress 
with the substitution Eq.~(\ref{add_drive}). At longer times, the form of the profiles
can not be given by a simple function anymore. In accordance with Fig.~\ref{wtsim},
in the case without residual stress the width of the profile is steadily increasing,
while in the presence of the residual stress, there is a regime where the edge advances faster than the
other end of the rim and the width changes non-monotonously. 

\begin{figure}
	\centering
	\includegraphics[width=0.45\textwidth]{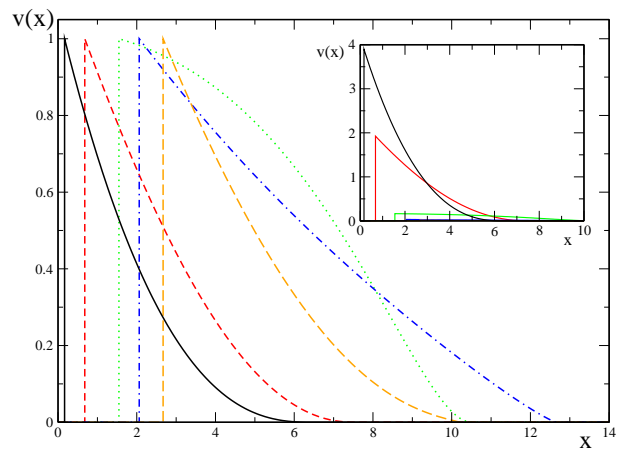}
	\caption{\label{vprof_ws}(Color online)
	The velocity  profiles $v(x)$, with an initial residual stress of $\sigma(t=0)=1$.
	The successive times are again as in the two previous figures
	and the inset shows the unrenormalized, rapidly decaying velocities. 
	}
\end{figure}

In the course of the dewetting process, 
the velocity profiles change more severely than the height profiles.
Fig.~\ref{vprof_wos} shows the velocity profiles inside the film for successive times
in the absence of initial residual stress, and
Fig.~\ref{vprof_ws} for the case of $\sigma(t=0)=1$.
In both figures the velocities have been renormalized to $1$ at the edge, 
in order to make the changes in the shape visible. The insets show the unrenormalized velocities 
that are rapidly decreasing in amplitude, as can be seen from Fig.~\ref{vtsim} for the (maximum)
value at the edge.
The short-time profiles are in accordance with the predictions of 
Eqs.~(\ref{vprofalpha}) and (\ref{add_drive}). The velocity as a function of
the distance from the edge, $x-L$, is rapidly decaying. However, when
the system is in the elastic regime ($\tau_0<t<\tau_1$), the
profiles are severely perturbed and become concave,
see the green dotted curves in both Fig.~\ref{vprof_wos} and Fig.~\ref{vprof_ws}. 
After the system has left the elastic regime and has again become viscous,
for $t>\tau_1$, the behavior is distinct for the two cases, see the blue (dash-dotted) and orange 
(long-dashed) curves:
without residual stress the shape of the profile becomes convex again with
similar (actually still slowly increasing) widths. 
With residual stress, the profiles also regain a convex shape, but 
first with a large characteristic width 
that then retracts. 

\begin{figure}
	\centering
	\includegraphics[width=0.45\textwidth]{fig/fig7}
	\caption{\label{vprof_wos}(Color online)
	The velocity profiles $v(x)$, without residual stress, at the 
	successive times as in Fig.~\ref{hprof}.
	The velocities at the edge have been normalized
	to one to make the changes in the profile visible.  
	The inset shows the actual profiles, where the edge velocity is rapidly decaying,
	cf. Fig.~\ref{vtsim}.
}
\end{figure}

The concave, rounded shape of the velocity profiles in the elastic regime can be
understood qualitatively from the governing equations.
In the elastic regime, 
Eq.~(\ref{const}) states that the dominating contribution to the stress 
is $\p_{t}\sigma\simeq G\p_{x} v$. If we assume linear friction for simplicity,
for the velocity we have $\zeta v=\p_x\left(h\sigma\right)$, leading to 
\beq\label{diffus_eq}
\p_t v\simeq \zeta^{-1}G \p_x(h_p\p_x v)\,.
\eeq
Here we have used the fact that the height is almost time-independent in the elastic plateau, 
see Fig.~\ref{htsim}, and has an almost stationary profile (in the frame moving with $L$) 
that we call $h_p(x)$.
Eq.~(\ref{diffus_eq}) is a generalized diffusion equation 
for the velocity field. 
The initial condition to think of is (upon entering the elastic regime) 
an exponential profile, cf. Eq.~(\ref{vlin}).
Additionally, there is a boundary condition at the edge, namely 
that $v(L)$ is decaying in time, cf. Fig.~\ref{vtsim}.
It is easy to convince oneself that indeed the exponential
evolves towards a concave, rounded shape in the course of time.
This happens the more severely the larger $G/\zeta$ is, 
and the faster $v(L)$ decays - which is the reason for the shape being
more severely perturbed in the case of residual stress, where $v(L)$
decays much faster, see Figs.~\ref{vprof_wos} and \ref{vprof_ws}.
The spatially dependent 'diffusion coefficient' $h_p(x)$, that decreases 
with the distance from the edge, is enhancing the rounding up close to the edge. 
The same general argument holds in case of nonlinear friction,
although it is less obvious since Eq.~(\ref{diffus_eq}) becomes a nonlinear diffusion equation.  

\begin{figure}
	\centering
	\includegraphics[width=0.45\textwidth]{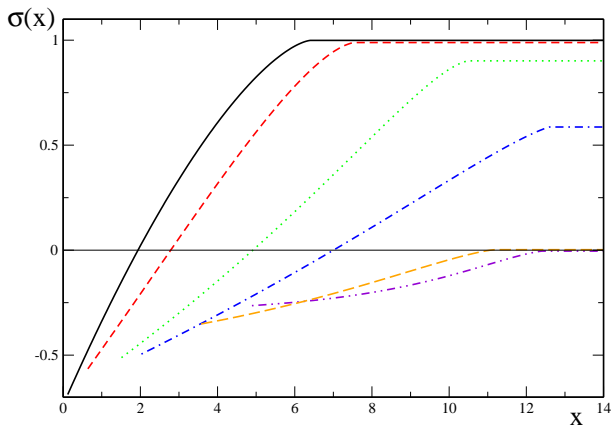}
	\caption{\label{sprof_ws}(Color online)
	The stress profiles $\sigma(x)$, with an initial residual stress of $\sigma(t=0)=1$.
	The successive times are: black (solid) $t=0.1$, red (dashed) $t=1$, 
	green (dotted) $t=10$, blue (dash-dotted) $t=50$, orange (long dashes) $t=500$, 
	violet (dash-double dot) $t=1000$.
	}
\end{figure}

The profiles for the stress inside the film are shown in Fig.~\ref{sprof_ws}, in the
presence of residual stress. The short time profile is in accordance 
with Eq.~(\ref{sprof_ana}). At the edge, the stress is determined by the boundary condition,
Eq.~(\ref{bound_ren}). 
The relaxation of the initial residual stress can be deduced
from the value at large $x$. Relaxation is complete after several relaxation times $\tau_1$.

%
\section{Energy balance}
\label{balance}
%

Since the work described in Refs.~\cite{FBW:1994,FBW:1997}, it has been
proven successful to describe the temporal behavior of dewetting by 
an energy balance that accounts for the relevant driving and 
dissipation mechanisms. This method has also been recently used to study the
dewetting in the present model by means of scaling laws \cite{Raphael:2006.1}. 
The energy balance can be derived from 
Eq.~(\ref{mech_n}) by multiplication by $v(x)$ and subsequent integration over space,
which yields
\beq\label{int_veq}
\frac{2-\al}{2}\int_L^\infty\hspace{-2mm} v(x)^{2-\al}dx
=2\int_L^\infty\hspace{-2mm} \p_x\Big(h(x)\sigma(x)\Big)v(x)dx\,.\,\,
\eeq
For brevity we have suppressed
the temporal dependence in the fields $v$, $h$ and $\sigma$, 
but this equation has to hold for all times $t>0$. 
The left hand side is the rescaled dissipation by friction. It is obviously
positive, thus also the right hand side has to be so.
Integration by parts of the right hand side, making use of the boundary conditions 
which imply
$\left[h\sigma v\right]_L^\infty=-h(L)\sigma(L)v(L)=\sqrt{2}v(L)$,
and rearrangement of terms results in
\beq\label{en_bal}
\sqrt{2}\,v(L)=\frac{2-\al}{2}\int_L^\infty\hspace{-2mm} v^{2-\al}dx
+2\int_L^\infty\hspace{-2mm} h\sigma\p_x vdx\,.\,\,
\eeq
This formula has a simple interpretation: the left hand side is the
work done by the rescaled driving force. In dimensional form it is proportional to $\aS v(L)$ and
thus has units of force/length times velocity. It naturally appeared as a boundary term.
On the right hand side, the first term is the dissipation by the (in general nonlinear) friction 
of the film with the substrate, and the second term is the height averaged 
(hence the weight factor $h(x)$)
dissipation inside the film, which is of the usual form $\sigma\p_x v$.

\subsection{Effect of residual stress}

In the presence of residual stress, we again write
$\sigma^{\rm tot}(x)=\sigma^{\rm v}(x)+\sigma_0$.
From Eq.~(\ref{int_veq}) we thus obtain 
\beq\label{en_bal2}
0&=&\hspace{-1mm}\frac{2-\al}{2}\hspace{-1mm}\int_L^\infty\hspace{-2mm} v^{2-\al}dx\nonumber\\
&&\hspace{-4mm}-\sqrt{2}\,v(L)
+2\hspace{-1mm}\int_L^\infty\hspace{-2mm} h\sigma^{\rm v}\p_x vdx
+2\sigma_0\hspace{-1mm}\int_L^\infty\hspace{-2mm} h\p_x vdx\,.\quad\,\,\,
\eeq
where we used that the residual stress is considered homogeneous here.
Both the second term (proportional to $-v(L)<0$) and
the last term are negative (since in general $\p_xv<0$, while $h>0$ and $\sigma_0>0$).
Thus these terms should be interpreted as the work done by the increased driving force
due to the residual stress \cite{comment2}.
The remaining terms are positive (for the third term one has to note that 
$\sigma_{\rm v}<0$ and $\p_xv<0$) 
and as before can be interpreted as the dissipation by friction
and the viscous dissipation inside the film, respectively.
In total, we get the balance
\beq\label{en_bal_s}
& &\sqrt{2}v(L)-2\sigma_0\hspace{-1mm}\int_L^\infty\hspace{-2mm} h\p_x vdx\nonumber\\
&=&\frac{2-\al}{2}\int_L^\infty\hspace{-2mm} v^{2-\al}dx
+2\int_L^\infty\hspace{-2mm} h\sigma^{\rm v}\p_x vdx\,.\,\,
\eeq
It is of the same form as Eq.~(\ref{en_bal}), but with the additional driving arising
from the residual stress. 
A similar relation was introduced empirically in Ref.~\cite{Raphael:2006.1}. 
Note that since the dynamics of the residual stress is governed by the constitutive relation,
Eq.~(\ref{const_n}), $\sigma_0$ in Eq.~(\ref{en_bal_s}) is time-dependent 
and relaxes like $\sigma_0(t)=\sigma_0(0) e^{-t/\tau_1}$.

%
\subsection{Numerical evaluation}
\label{num_balance}
%

It has been shown for viscous fluids \cite{FBW:1997} that
if a rim has already been formed on a slippery substrate, 
the viscous dissipation should be negligible as compared to the
dissipation by friction with the substrate. 
Recently it has been shown \cite{Vilmin_slippery},
that for a linear friction law and in the initial (viscous) 
regime of dewetting, both mechanisms contribute equally to the dissipation. 
What happens at intermediate times, in the viscoelastic regime, 
and what are the effects of nonlinear friction and residual stress is explored in this section. 

\begin{figure}
	\centering
	\includegraphics[width=.45\textwidth]{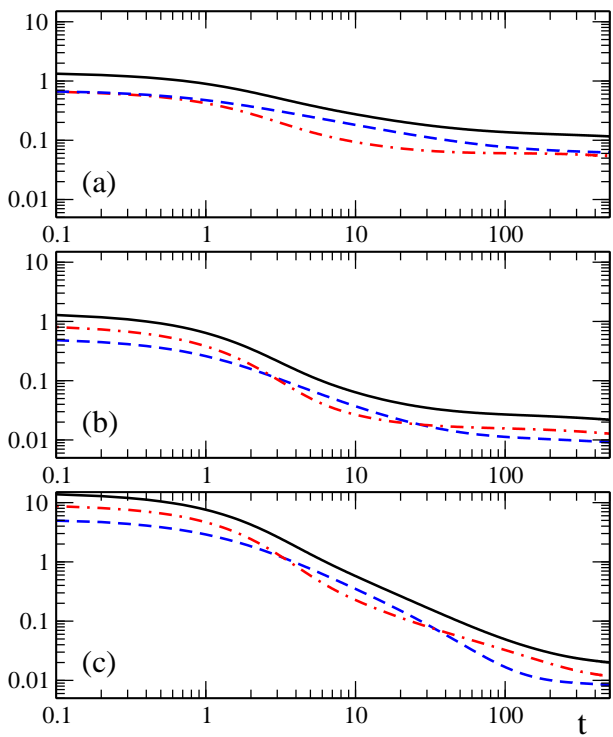}
	\caption{\label{dissip}(Color online)
	Numerical evaluations of the energy balance, Eq.~(\ref{en_bal_s}).
	The black solid curves show the work done by the driving force, $\Delta W$, the 
	red dash-dotted curves the dissipation in the film, $T\dot S_{v}$, and the
	blue dashed curves the dissipation by friction with the substrate, $T\dot S_{f}$.
	Panels (a) and (b) correspond to the case with linear and nonlinear friction law, 
	respectively, in the absence of residual stress, while panel (c) shows the case of
	nonlinear friction and an initial residual stress $\sigma(t=0)=1$.
	}
\end{figure}

Knowing all the relevant fields from the numerical solution, 
using Eq.~(\ref{en_bal})  (or Eq.~(\ref{en_bal_s})
in the presence of residual stress)
we have access to the work done by the driving force, 
$\Delta W=\sqrt{2}v(L)-2\sigma_0\hspace{-1mm}\int_L^\infty\hspace{-2mm} h\p_x vdx$,
the dissipation by friction $T\dot S_{f}=\frac{2-\al}{2}\int_L^\infty\hspace{-2mm} v^{2-\al}dx$ 
and the dissipation inside the film $T\dot S_{v}=2\int_L^\infty\hspace{-2mm} h\sigma^{\rm v}\p_x vdx$. 
These three quantities are shown in Fig.~\ref{dissip}:
the upper panel (a) shows the case with linear friction ($\al=0$) and without residual stress.
The middle (b) and lower (c) panel have been obtained with nonlinear friction, 
without residual stress and with $\sigma(t=0)=1$, respectively.
In all panels the black (solid), red (dash-dotted) and blue (dashed) lines correspond to $\Delta W$, $T\dot S_{v}$ and $T\dot S_{f}$
respectively.
As it should, the sum of the blue and the red curves, i.e. the total dissipation, equals 
the black curve, which is the work done by the driving force, confirming that the
energy balance is respected.

At short times and in the case of a linear friction law, Fig.~\ref{dissip}(a) shows that 
the two dissipation mechanisms contribute equally, 
in accordance with Ref.~\cite{Vilmin_slippery}.
This can also be seen by directly evaluating Eq.~(\ref{en_bal}) with the
analytical short-time solutions obtained in section~\ref{anashort}.
In the elastic regime, for $\tau_0<t<\tau_1$, the
dissipation by friction is larger than the dissipation in the film. Since the system
is rather elastic than viscous, the dissipation in the film is reduced, until it becomes again
comparable to the frictional dissipation for $t>\tau_1$.
Thus the dissipation by friction is always more or at least equally  
important than the dissipation inside the film.
This is in contrast to the case with nonlinear friction displayed in Fig.~\ref{dissip}(b) and (c). 
In this case, at short times the viscous dissipation is dominating, since the sublinear friction law
reduces the friction. In the elastic regime, the frictional dissipation becomes dominating,
by the same argument than in case of linear friction.
For $t>\tau_1$, where the system becomes viscous again, the dissipation in the film 
is dominating, especially in the case with residual stress shown in Fig.~\ref{dissip}(c).
This change in the importance of the two dissipation mechanisms is clearly a
combined effect of both viscoelasticity and nonlinear friction.

\section{Discussion}
\label{Disc}

We are now able to discuss the effects of the viscoelasticity, the nonlinear friction and
the initial residual stress and their mutual interplay in some detail.
The viscoelasticity clearly is responsible for the 
plateau in the height of the rim, cf. Fig.~\ref{htsim}, that is also experimentally observed. 
Concerning the flow profiles, it leads to a rounded concave
form of the velocity profile, cf. Fig.~\ref{vprof_wos}
that can be explained by the elasticity and the decrease in the driving force
due to the rim build-up. 
When the system enters the second viscous regime, around $\tau_1$,
the elastic energy stored in the flow profiles is dissipated in the film, leading
to a relative increase of the viscous dissipation with respect to frictional dissipation.
 In case of a linear friction law
with the substrate both dissipation mechanisms, the one by 
friction on the substrate which was dominating in the elastic regime and the one by viscous
dissipation inside the film, become approximately equally important 
for $t>\tau_1$, cf. Fig.~\ref{dissip}(a). 
In case of nonlinear friction the dissipation in the film becomes even more important than the
dissipation by friction, see Fig.~\ref{dissip}(b),(c). Though, on even longer time scales
(not captured by the model presented here) where
the mature regime is reached and the profile becomes a half-cylinder with a stationary
flow profile while the rim is still growing, the dissipation by friction will be dominating again
\cite{FBW:1997}.

The nonlinear friction renders all the profiles steeper, i.e. those for the height of the film
as well as those for the velocity and the stress inside the film.  Together with the viscoelasticity,
the nonlinear friction leads to two crossovers concerning the importance of the two dissipation
mechanisms: while in the two viscous regimes (for $t<\tau_0$ and $t>\tau_1$) the viscous
dissipation in the film is more important, since the friction is reduced due to the
sublinear friction law, in between $\tau_0$ and $\tau_1$ 
the dissipation by friction on the substrate is
dominating. This is due to the fact that the polymer film is predominantly elastic
in this regime and viscous dissipation is reduced.

The residual stress leads, as expected, to an increase of the driving force, cf. Eq.~(\ref{add_drive}),
consequently resulting in a faster initial dynamics. 
Concerning the velocity profiles, the faster decrease of the driving force due
to the relaxation of the initial stress  amplifies the roundup of the profiles, cf.
Figs.~\ref{vprof_wos} and \ref{vprof_ws}. Hence, when the system changes
from elastic to viscous around $\tau_1$, the dissipation inside the film
is larger than without stress. This indicates that the occurrence of the maximum
in the rim width is due to the faster decrease in the driving by stress relaxation, 
coupled to the increased importance of dissipation inside the film that was caused
by both the viscoelasticity and the nonlinearity in the friction. 

\section{Conclusions and perspective}
\label{Concl}

In conclusion, we have investigated numerically the temporal evolution of a dewetting thin polymer
film under homogenous lateral stress. The main properties of the polymeric system, 
namely the viscoelasticity of the film, the friction with the substrate
and the residual stress have been shown to interplay in the course of dewetted time,
leading to complex behavior. 
Numerical evaluations of the energy balance, derived from the
underlying model equations, revealed that the dissipation during the dewetting
is more complex than expected. The viscous dissipation in the
film and the frictional dissipation at the substrate have different time dependences
and, in case of a nonlinear friction law, the friction is not always the most important
dissipation mechanism. 
Especially, the occurrence of a maximum in the rim width, as observed 
experimentally in Refs.~\cite{Reiter:2005,Reiter:2007.2},
could be traced back to the combined effect of a more rapid decrease in the driving force
due to relaxation of stress and the viscous dissipation of the 
elastic energy stored in the flow profile, which is the dissipation mechanism
with the largest contribution
at the time scale of the elastic-viscous crossover due to the nonlinearity of the friction.

Our numerical treatment also sheds new light on simple approaches based
on scaling laws put forward previously \cite{FBW:1997,Vilmin_nlfric,Raphael:2006.1}.
There one either assumes that the dissipation by friction 
dominates over the viscous dissipation, or that both mechanisms have the same temporal dependence.
Additionally, in these approaches one has to assume a simple form of volume conservation, namely 
$h_0 L(t)=C(H(t)-h_0)W(t)$, to connect 
the rim width $W$ with the dewetted length $L$ and the height $H$ at the edge, with a fixed 
constant $C$ depending on the shape of the rim.  
The complex behavior concerning the dissipation and the non self-similarity
of the profiles indicates that these simple scaling arguments 
should be revisited. Indeed, if dissipation by friction would 
be the only important dissipation mechanism, 
by balancing the driving with the frictional dissipation, 
a decrease in the width of the rim
would result in a speed up of the dewetting velocity, which is never observed. 

The inclusion of a residual stress in the model 
proved to be crucial for the occurrence
of the maximum in the rim width. This maximum has been already used to extract
relevant information from experiments, namely the exponent of the nonlinear
friction law \cite{Vilmin_nlfric}. The inclusion of residual stresses has also
been put forward to analyze effects of film ageing: 
The dynamics of the dewetting has been shown to be
severely influenced by the age of the sample, 
both on solid and liquid substrate \cite{Reiter:2005,Fretigny}. 

Clearly it would be desirable to extract more information on these stresses
both from the technological point of view of film stability
and for fundamental reasons to better understand 
the ageing of a confined glassy polymer film by means of dewetting experiments.
Less is known so far on how these stresses look like - what are the 
non-equilibrium configurations and the relaxation dynamics
of the polymer chains confined in such a thin film?
There is no special reason, a priori, for the residual stress 
to relax with the same time scale that governs the long-time flow behavior 
of the polymeric liquid - as we have assumed here for simplicity. 
Indeed, it has been recently observed experimentally \cite{Reiter:2007.2}
that there are (at least) two time scales:
the characteristic time of the maximum in the rim width, which 
does not scale like reptation with molecular weight, and the long-time crossover 
in the dewetting velocity, which does. These issues will be investigated in 
a forthcoming publication. 

We thank G\"unter Reiter and Pascal Damman for very stimulating discussions
and Michael Schindler for valuable comments on the manuscript.
F.Z. acknowledges financial support by 
the European Community's ''Marie-Curie Actions'' under 
contract MRTN-CT-2004-504052 (POLYFILM) and by the ESPCI (Chaire Joliot).



\begin{thebibliography}{37}
\bibitem{comment}
There is a contribution from the pressure to the stress, 
$\sigma=\eta\p_x v-P$. With $P=-\eta\p_x v$ from the
lubrication approximation \cite{Raphael:2006.1},
this effectively doubles the viscosity. To be consistent,
$\eta_0$ from the Jeffrey model has to be identified with $2\eta$, 
where $\eta$ is the 'actual' fluid viscosity.
\bibitem{comment2}
The term $-2\sigma_0\hspace{-1mm}\int_L^\infty\hspace{-1mm} h\p_x vdx$ in Eq.~(\ref{en_bal_s})
is not a pure boundary term. It becomes proportional to $v(L)$ only for $h=$const. 
Thus in general it contains also a small part of (frictional) dissipation.
\end{thebibliography}

\end{document}